\documentclass[aps,prd,preprint,preprintnumbers]{revtex4}

\usepackage{graphicx}
\usepackage[catalan, USenglish]{babel}
\usepackage[usenames,dvipsnames]{color}
\usepackage[utf8]{inputenc}
\usepackage{amsmath}
\usepackage{amsfonts}
\usepackage{amssymb}
\usepackage{enumitem}   
\usepackage{setspace}
\usepackage{multirow}

\newcommand{\AddrAHEP}{AHEP Group, Institut de F\'{i}sica Corpuscular --
  C.S.I.C./Universitat de Val\`{e}ncia, Parc Cientific de Paterna.\\
  C/Catedr\'atico Jos\'e Beltr\'an, 2 E-46980 Paterna (Val\`{e}ncia) - SPAIN}

\newcommand{\Cinvestav}{Departamento de F\'{\i}sica, Centro de
  Investigaci{\'o}n y de Estudios Avanzados del IPN\\ Apdo. Postal
  14-740 07000 Mexico, DF, Mexico}
  
\newcommand{\AddrUNAM}{Instituto de F\'{\i}sica, Universidad Nacional
Aut\'onoma de M\'exico, A.P. 20-364, Ciudad de M\'exico 01000, M\'exico.}
\begin{document}

\title{Neutrino counting experiments and non-unitarity from LEP and future 
experiments}
\author{F. J. Escrihuela~$^1$}\email{franesfe@alumni.uv.es} 
\author{L. J. Flores~$^{2,3}$}\email{jflores@fis.cinvestav.mx}
\author{O. G. Miranda~$^2$}\email{omr@fis.cinvestav.mx}

\affiliation{$^1$~\AddrAHEP}
\affiliation{$^2$~\Cinvestav}
\affiliation{$^3$~\AddrUNAM}

\begin{abstract}
  Non-unitarity of the neutrino mixing matrix is expected in many
  scenarios with physics beyond the Standard Model. Motivated by the
  search for deviations from unitary, we study two neutrino
  counting observables: the neutrino-antineutrino gamma process and
  the invisible $Z$ boson decay into neutrinos. We report on new
  constraints for non-unitarity coming from the first of this
  observables.  We study the potential constraints that future
  collider experiments will give from the invisible decay of the Z
  boson, that will be measured with improved precision.
\end{abstract}

\maketitle

\section{Introduction}
Particle physics is currently in an era of great progress, with new
experiments~\cite{Baer:2013cma,Gomez-Ceballos:2013zzn,CEPCStudyGroup:2018ghi,CEPC-SPPCStudyGroup:2015csa,CLIC:2016zwp}
envisaged for the future.  The existence of neutrino oscillations, as
well as the discovery of the Higgs Boson are the main motivations for
the development of new experiments that will measure the standard
physics parameters with unprecedented precision, while also searching for
new physics.

In the Standard Model picture, there are three active light neutrinos
with an interaction governed by the $SU(2)_L\otimes U(1)_Y$
electroweak symmetry~\cite{Schechter:1980gr}. The neutrino mixing in
this case is described by an unitary $3 \times 3$ matrix. If more
(heavy) neutrino states exist, the corresponding mixing matrix will be
bigger and it will have, at some level, a deviation from
unitarity. Such picture has been studied since long time
ago~\cite{Nardi:1994iv,Antusch:2006vwa, Smirnov:2006bu,Dev:2009aw, Ohlsson:2010ca,
    Akhmedov:2013hec, deGouvea:2015euy, Antusch:2016brq} and, more
recently, a description in terms of a triangular parameterization has
been
discussed~\cite{Escrihuela:2015wra,Escrihuela:2016ube,Miranda:2016ptb,Blennow:2016jkn}.

In the presence of such a non-unitary (NU) mixing, neutrino counting
experiments at high energies will differ from the Standard Model
prediction~\cite{GonzalezGarcia:1990fb}. This is the case of
the invisible decay width of the $Z$ boson~\cite{ALEPH:2005ab,
  Schael:2013ita} and also of the $\nu\bar{\nu}\gamma$
measurements~\cite{Barate:1997ue, Barate:1998ci,
	Heister:2002ut, Abreu:2000vk, Acciarri:1997dq, Acciarri:1998hb,
	Acciarri:1999kp, Ackerstaff:1997ze, Abbiendi:1998yu,
	Abbiendi:2000hh, Hirsch:2002uv}. As far as we know, no constraints
on non-unitarity have been reported from the $\nu\bar{\nu}\gamma$
process. On the opposite side, the current measurement of the invisible
decay of the $Z$ boson lies two standard deviations below the Standard
Model prediction, a measurement that has already been studied with
detail~\cite{Carena:2003aj}. On the other hand, different proposals
for the future generation of collider experiments are currently under
development~\cite{Fan:2014vta}, such as ILC~\cite{Baak:2013fwa, 
	Baer:2013cma, Banerjee:2015gca},
FCC-ee~\cite{Gomez-Ceballos:2013zzn, Blondel:2014bra}, and CEPC~\cite{CEPCStudyGroup:2018ghi,CEPC-SPPCStudyGroup:2015csa,
    Liao:2017jiz, Liang:2018mst}. These proposals will be running at
the very high energy regime, searching for new physics and measuring
the Standard Model parameters in a different energy scale. They will
also test physics at relatively lower energies, in order to improve
the measurements on already known observables. In particular, it is
expected that the invisible $Z$ decay width will be measured with
improved precision, if compared to the current reported measurement by
LEP~\cite{ALEPH:2005ab,Schael:2013ita}.

In this work we study the constraints arising from the neutrino
counting experiments around the $Z$ peak, specifically using
data from $\nu\bar{\nu}\gamma$ measurement. We also analyze the invisible $Z$ decay to have
a complete scenario in the same framework and study the potential of
future neutrino counting experiments in the same energy regime to
constraint the non-unitary parameters, and compare these perspectives
with the current constraints. We will show that the perspectives in
these future experiments are very promising. 

In section II, we will start the discussion by describing the
non-unitarity formalism that we will use. Then, in section III we
present the analysis used to obtain constraints on the non-unitary
parameters, as well as the found results and perspectives for future
experiments. Finally, in section IV we present our conclusions.

\section{Non-unitarity, invisible $Z$ decay and $\nu\bar{\nu}\gamma$} 

Non-unitarity has been subject to study for a long
time~\cite{Schechter:1980gr, Gronau:1984ct, Nardi:1994iv,
  Atre:2009rg}. Recent constraints can be found
elsewhere~\cite{Escrihuela:2016ube,Fernandez-Martinez:2016lgt}, either
considering only the restrictions coming from neutrino experiments, or
including the ones from charged leptons.  In both cases it is useful
to consider the mixing matrix as describing the transformation of
three light neutrinos and $n-3$ neutral heavy leptons. In this way,
one can see the $U^{n\times n}$ matrix as the combination of four
submatrices~\cite{Hettmansperger:2011bt}
\begin{equation}
U^{n\times n}=\left(\begin{array}{cc} N & S\\
V & T
\end{array}\right)\label{eq:ULindner_C1} ,
\end{equation}
with $N$ a $3\times3$ submatrix in the light neutrino sector, and $S$ 
the $3\times (n-3)$
submatrix that describes the mixing of the extra heavy isosinglet
states.

One useful way to
parameterize the non-unitarity of the mixing matrix $N$ is the
triangular parameterization~\cite{Escrihuela:2015wra}

\begin{equation}
 N
 =
 N^{NP} U
 =
 \left(
 \begin{array}{ccc} 
 \alpha_{11} & 0 & 0\\
 \alpha_{21} & \alpha_{22} & 0\\
 \alpha_{31} & \alpha_{32} & \alpha_{33}
 \end{array}
 \right) U \,,
 \label{eq:NU}
\end{equation}
where $U$ is the unitary PMNS mixing matrix for the standard $3\times 3$
case and $N^{NP}$ parameterizes the deviations from unitarity. In this way, we can encode
all the parameters of the general
description~\cite{Schechter:1980gr,Rodejohann:2011vc}, for an
arbitrary number of additional neutrino states, in a compact notation.
In this general framework, we can describe the non-unitary
phenomenology by using the three real parameters
$\alpha_{11},\alpha_{22}$, and $\alpha_{33}$ (all of them close to
one) plus other three complex parameters
$\alpha_{21},\alpha_{31},\alpha_{32}$ that contains extra CP violating
phases and whose magnitude is small.

In what follows we will discuss two neutrino counting observables in the 
context of this triangular parameterization. 

\subsection{The invisible $Z$ decay}

In the standard unitary limit, the branching for the invisible $Z$ decay 
into neutrinos will be given by~\cite{Akhmedov:1999uz, Tanabashi:2018oca},
\begin{equation}
\Gamma_{inv}= N_{\nu}\Gamma_{\nu \bar{\nu}}
\end{equation}
with $N_\nu$ the effective number of neutrino families and~\cite{Tanabashi:2018oca} 
\begin{equation}
\Gamma_{\nu
  \bar{\nu}}=\frac{G_F M_Z^3}{12 \sqrt{2}\pi}.
\end{equation}
Experimentally, the ratio $\Gamma_{inv}/\Gamma_{\ell \bar{\ell}}$ has
been measured with greater experimental precision than $\Gamma_{inv}$
alone~\cite{Tanabashi:2018oca, ALEPH:2005ab}. Therefore, the number of light active neutrinos can be estimated
from this relation, that in the Standard Model is  given by~\cite{ALEPH:2005ab}
\begin{equation} \label{ratioExp}
R^0_{inv} \equiv \frac{\Gamma_{inv}}{\Gamma_{\ell \bar{\ell}}} = N_\nu \left(\frac{\Gamma_{\nu\bar{\nu}}}{\Gamma_{\ell \bar{\ell}}}\right)_{SM} ,
\end{equation}
with $N_\nu = 3$. Here, the decay rate for the $Z$ boson into charged
leptons is given by~\cite{Tanabashi:2018oca}
\begin{equation}\label{gammaLep}
  \Gamma_{\ell \bar{\ell}} = \frac{G_F M_Z^3\left({g_V^\ell}^2 + {g_A^\ell}^2\right)}{6 \sqrt{2}\pi},
\end{equation}
where $g_V^\ell$ and $g_A^\ell$ are the vector and axial coupling for
a charged lepton $\ell$:
\begin{equation}
   g_V^\ell = T_\ell - 2Q_\ell \sin^2\theta_W\, ,
\nonumber
\end{equation}
\begin{equation}
   g_A^\ell = T_\ell .
\nonumber
\end{equation}

When we consider the non-unitarity formalism, applied to the invisible
decay rate of the $Z$ boson, we will find that the contribution of the
three active neutrino flavors will be given by~\cite{Tanabashi:2018oca}  
\begin{equation}\label{rateNeutrino3}
	\Gamma_{inv} = \frac{G_F M_Z^3 \sum_{i,j}|(N^\dagger N)_{ij}|^2}{12 \sqrt{2}\pi},
\end{equation}
that can also be expressed as 
\begin{equation} \label{rateNeutrino4}
	\Gamma_{inv} = \frac{G_F M_Z^3 \sum_{\alpha,\beta}|(NN^\dagger)_{\alpha\beta}|^2}{12 \sqrt{2}\pi} .
\end{equation}
Comparing this expression with the unitary case discussed before, we
can define for the non-unitary case
\begin{equation} \label{rateNeutrino}
	N_{\nu} = \sum_{\alpha, \beta}|(NN^\dagger)_{\alpha\beta}|^2 .
\end{equation}

It is important to notice that the theoretical expression for the
decay rate will be affected by non-unitarity with several corrections.
However, we must notice that there is another correction due to the
definition of $G_F$.
In order to introduce this correction,  we can
write the equivalent expression to Eq.~(\ref{ratioExp}) for the non-unitary case. For this purpose, we start by considering that, from
muon decay, a non-unitary mixing will affect the value of the Fermi
constant to be~\cite{Nardi:1994iv, Langacker:1988ur, Atre:2009rg} 
\begin{equation}
  \label{GfNU}
  G_{F} = \frac {G_{\mu}}{\sqrt{\sum_{ij}|N_{\mu i}|^2|N_{e j}|^2}} =
  \frac
  {G_{\mu}}{\sqrt{\alpha_{11}^{2}(\alpha_{22}^{2}+|\alpha_{21}|^{2})}}
  \, .
\end{equation}

This correction cancels out in the ratio, $R_{inv}^0$, but can
propagate to other observables, such as the weak mixing angle~\cite{Fernandez-Martinez:2016lgt}
\begin{equation}\label{weakAngle} 
\sin^2\theta_W = \frac{1}{2}\left(1 - \sqrt{1 - \frac{2\sqrt{2}\alpha\pi}{G_\mu M_Z^2}\sqrt{\alpha_{11}^{2}(\alpha_{22}^{2}+|\alpha_{21}|^{2})}} \right). 
\end{equation}
		
From Eqs.~(\ref{rateNeutrino}) and~(\ref{gammaLep}), we can get an
expression for the ratio in the non-unitary case:
\begin{equation}
R^0_{inv} = \frac{\sum_{\alpha,\beta}|(NN^\dagger)_{\alpha\beta}|^2} {2({g_V^\ell}^2 + {g_A^\ell}^2)}.
\end{equation}

Let us notice that the deviation from unitarity, introduced by the
parameters $\alpha_{ij}$, appears explicitly in the numerator through
$|(NN^\dagger)_{\alpha\beta}|^2$, but also implicitly in the
denominator via $g_V^\ell$, because it contains the expression for the weak
mixing given in Eq.~\eqref{weakAngle}. The explicit form for the
numerator in the previous formula will be
\begin{eqnarray}\label{eq:NN}
\sum_{\alpha,\beta}|(NN^\dagger)_{\alpha\beta}|^2 &=& \alpha_{11}^4 + \alpha_{22}^4 + \alpha_{33}^4 + |\alpha_{21}|^4 + |\alpha_{31}|^4 +|\alpha_{32}|^4 \nonumber \\
&+& 2\alpha_{22}^2|\alpha_{21}|^2 + 2\alpha_{33}^2(|\alpha_{31}|^2 +
|\alpha_{32}|^2) + 2|\alpha_{31}|^2|\alpha_{32}|^2 \, 
\phantom{\sum_{\alpha, \beta}} \\
&+& 2\alpha_{11}^2(|\alpha_{21}|^2 + |\alpha_{31}|^2) + 2|\alpha_{21}\alpha_{31}^* + \alpha_{22}\alpha_{32}^*|^2. \nonumber 
\end{eqnarray}
If we neglect terms including third order or higher on off-diagonal
parameters ($\alpha_{ij} \; i \neq j$), we obtain the following
reduced expression:
\begin{equation}
\sum_{\alpha,\beta}|(NN^\dagger)_{\alpha\beta}|^2 = \alpha_{11}^4 +
\alpha_{22}^4 + \alpha_{33}^4 +2\alpha_{11}^2(|\alpha_{21}|^2 + |\alpha_{31}|^2)+2\alpha_{22}^2(|\alpha_{21}|^2 +|\alpha_{32}|^2) +
2\alpha_{33}^2(|\alpha_{31}|^2 + |\alpha_{32}|^2) \, .
\phantom{\sum_{\alpha, \beta}} \nonumber 
\end{equation}
Different constraints for the $\alpha_{ij}$ parameters show that $N^{NP}$
is close to an identity matrix. 

Besides, the precision of the measurements under consideration makes
necessary to introduce radiative corrections. In the $\overline{MS}$
scheme, the weak mixing angle takes the form~\cite{Tanabashi:2018oca}
\begin{equation} \label{sin2_corr_rad}
   \hat{s}_{Z} = \frac{A_{0}}{M_{W}(1-\Delta \hat{r}_{W})^{1/2}}   , 
\end{equation}
where, in the non-unitary case,  $A_{0}$ is given by
\begin{equation} \label{A0}
A_{0} = \left(\frac{\pi \alpha}{\sqrt{2} G_{F}}\right)^{1/2} = \left(\frac{\pi \alpha \sqrt{\alpha_{11}^{2}(\alpha_{22}^{2}+|\alpha_{21}|^{2})}}{\sqrt{2} G_{\mu}}\right)^{1/2} ,	
\end{equation}
$\Delta \hat{r}_{W}$ introduces the radiative corrections, and $M_W$ is
the mass of the $W$ boson. According to PDG~\cite{Tanabashi:2018oca},
the values of the relevant parameters are:
\begin{eqnarray}
M_{W} &=& 80.379 \pm 0.012\; GeV/c^{2} \, , \\
\Delta \hat{r}_{W} &=& 0.06916 \pm 0.00008 \, , \\
 \alpha &=& (7.2973525664 \pm 0.0000000017) \times 10^{-3} \, , \\
G_{\mu} &=& (1.1663787 \pm 0.0000006) \times 10^{-5} \; GeV^{2} \\
{\hat{s}_{Z}}^{2} &=& 0.23122 \pm 0.00003 \, .
\label{rad_corrvalues_pdg}
 \end{eqnarray}
For measurements at energies around the $Z$ peak it is common to use
the effective weak mixing angle ${\bar{s}_{l}}^2$ instead of the
$\overline{MS}$ scheme; both quantities are related
through~\cite{Tanabashi:2018oca} ${\bar{s}_{l}}^{2} = {\hat{s}_{Z}}^{2} +
0.00032 $.

Now we can turn now our attention to the comparison
with the experimental results to obtain constraints and future
perspectives for the NU parameters. However, before entering into this
discussion we will also discuss another neutrino counting observable.

\subsection{The process $e^-e^+ \to \nu\bar{\nu}\gamma$}
Another process that was also measured at LEP, and allows for a
neutrino counting, is the single photon production with a
neutrino-antineutrino pair~\cite{Barate:1997ue, Barate:1998ci,
    Heister:2002ut, Abreu:2000vk, Acciarri:1997dq, Acciarri:1998hb,
    Acciarri:1999kp, Ackerstaff:1997ze, Abbiendi:1998yu,
    Abbiendi:2000hh, Hirsch:2002uv}. In this subsection we compute
the expression for this observable in the NU case.

The differential cross section for the single photon production from
electron-positron annihilation, $e^+e^- \to \nu\bar{\nu}\gamma$, can be written in terms of the radiator
function $H(x,y;s)$ and the ``reduced'' cross section for
the process $e^+e^- \to \nu\bar{\nu}$, $\sigma_0$, 
as~\cite{Nicrosini:1988hw, Barranco:2007ej,Berezhiani:2001rs}:
\begin{equation}\label{eq:differential}
\frac{d^2\sigma}{dx\,dy}= H(x,y;s)\, \sigma_0(s(1 -x)). 
\end{equation}
The radiator function is defined by 
\begin{equation}
H(x,y;s) = \frac{2 \alpha}{\pi} \frac{\left[(1-\frac{1}{2}x)^2 + \frac{1}{4}x^2y^2\right]}{x(1-y^2)},
\end{equation}
with
\begin{equation}
x = 2 E_\gamma / \sqrt{s}, \quad y = \cos\theta_\gamma, 
\end{equation}
and $\sigma_0$, the ``reduced'' cross section for the process $e^+e^-
\to \nu\bar{\nu}$ is given by
\begin{align}\label{eq:reducedCrossSec}
\sigma_0(s) &= \sigma_W(s) + \sigma_Z(s) + \sigma_{W-Z}(s), \nonumber \\
\sigma_0(s) &= \frac{G_F^2 s}{12\pi} \left[2 + \frac{N_\nu (g_V^2 + g_A^2)}{(1 - s / M_Z^2)^2 + \Gamma_Z^2 / M_Z^2} + \frac{2 (g_V + g_A) (1 - s / M_Z^2)}{(1 - s / M_Z^2)^2 + \Gamma_Z^2 / M_Z^2}\right] .
\end{align}
	
\begin{figure}[h]
	\begin{minipage}{0.05\textwidth}
		(a)
	\end{minipage}
	\begin{minipage}{0.3\textwidth}
		\includegraphics[scale=0.6]{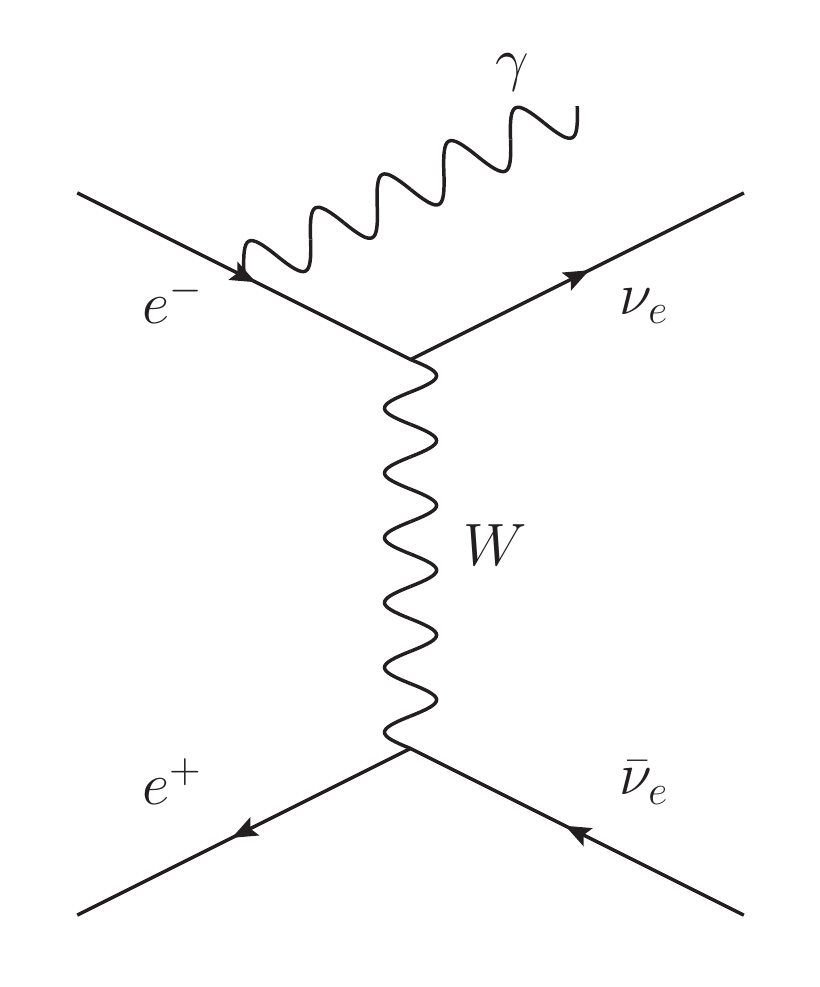}
	\end{minipage}
	\begin{minipage}{0.3\textwidth}\vspace{1.3cm}
		\includegraphics[scale=0.6]{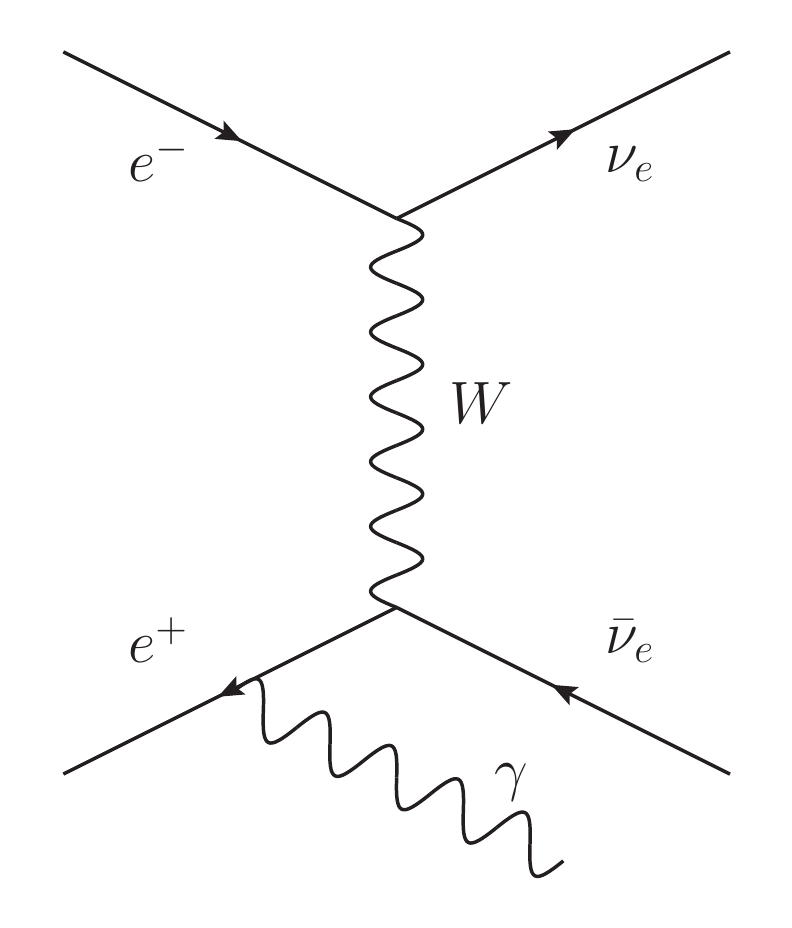}
	\end{minipage}
	\begin{minipage}{0.3\textwidth}
		\includegraphics[scale=0.6]{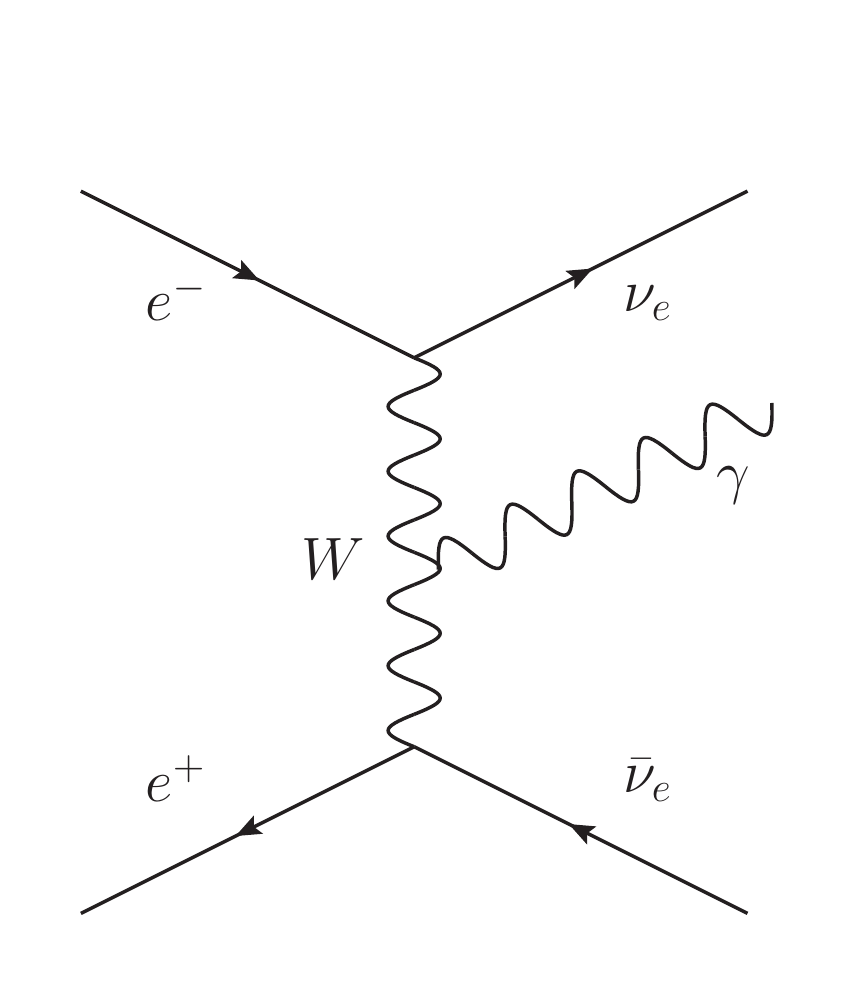}
	\end{minipage}
	\begin{minipage}{0.05\textwidth}
		(b)
	\end{minipage}
	\begin{minipage}{0.4\textwidth}
		\includegraphics[scale=0.6]{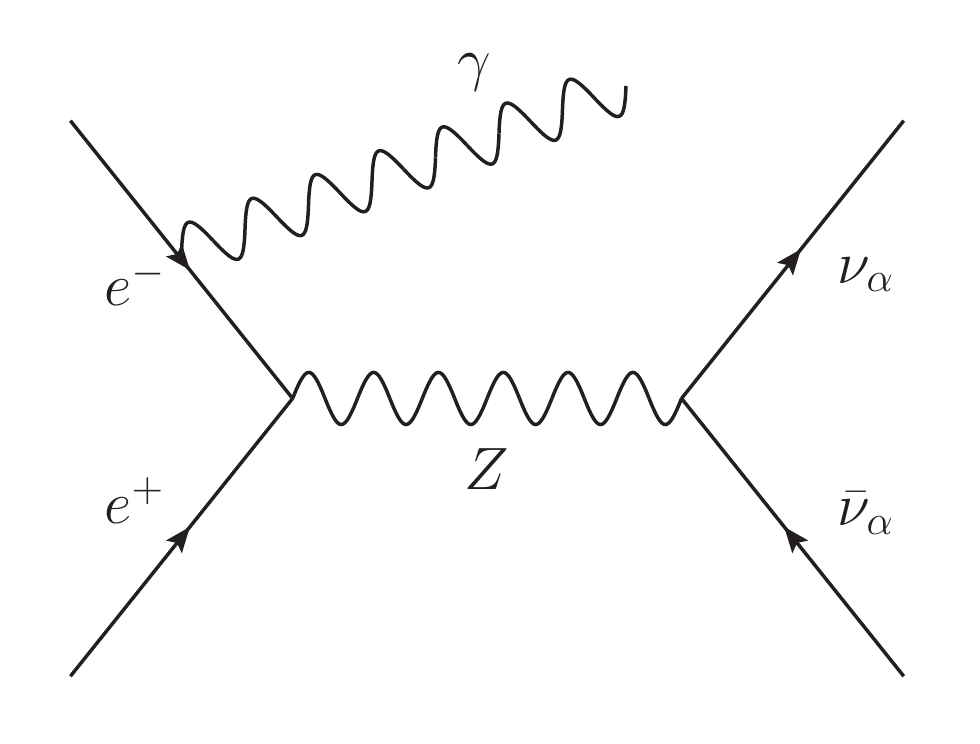}
	\end{minipage}
	\begin{minipage}{0.4\textwidth} \vspace{0.6cm}
		\includegraphics[scale=0.6]{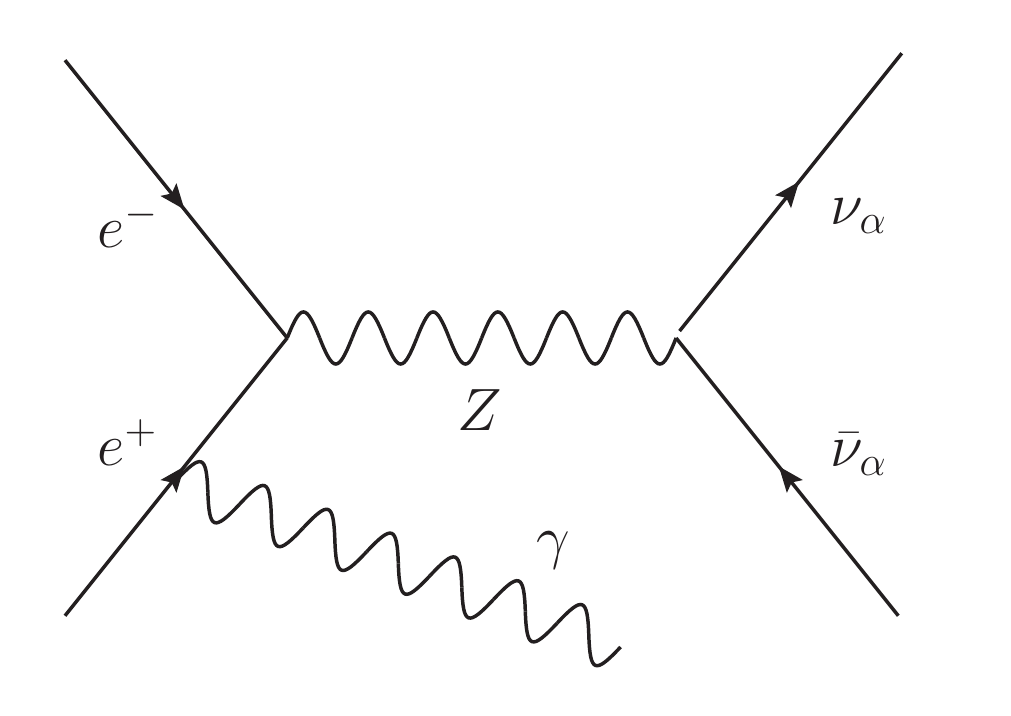}
	\end{minipage}
	\caption{Contributions to the $e^-e^+ \to \nu\bar{\nu}\gamma$ process at tree level, from the $W$ (a) and $Z$ (b) bosons.}
	\label{fig:diagrams}
\end{figure}

The three terms in Eq.~\eqref{eq:reducedCrossSec} come from the
contribution of the $W$, the $Z$ boson, and their interference, as can be seen in the 
Feynman diagrams in Fig.~\ref{fig:diagrams}.

For energies above the $Z$ resonance, finite distance effects on the
$W$ propagator need to be considered. These effects are taken into
account by the following substitution~\cite{Hirsch:2002uv, Barranco:2007ej,Berezhiani:2001rs}:
\begin{align}\label{eq:reducedCrossSec2}
\sigma_W(s) \; &\to \; \sigma_W(s) F_W(s/M_W^2), \nonumber \\
\sigma_{W-Z}(s) \; &\to \; \sigma_{W-Z}(s) F_{W-Z}(s/M_W^2),
\end{align}
where
\begin{align}
F_W(z) &= \frac{3}{z^3}\left[-2(z+1)\log(z+1) + z(z+2)\right], \nonumber \\
F_{W-Z}(z) &= \frac{3}{z^3}\left[(z+1)^2\log(z+1) - z(\tfrac{3}{2}z+1)\right].
\end{align}

It is important to notice that the expression in Eq.~\eqref{eq:reducedCrossSec}, including
the corrections from Eq.~\eqref{eq:reducedCrossSec2}, is equivalent to the widely used expression:
\begin{align}
\sigma_0(s) & = \frac{N_\nu G_F^2}{6\pi}M_Z^4 (g_R^2 + g_L^2) \frac{s}{\left[\left(s - M_Z^2\right)^2 + \left(M_Z \Gamma_Z\right)^2 \right]} \nonumber \\
& + \frac{G_F^2}{\pi} M_W^2 \Bigg\{ \frac{s + 2 M_W^2}{2s} - \frac{M_W^2}{s}\left(\frac{s + M_W^2}{s}\right) \log\left(\frac{s + M_W^2}{M_W^2}\right) \nonumber \\
& - g_L \frac{M_Z^2(s - M_Z^2)}{\left(s - M_Z^2\right)^2 + \left(M_Z \Gamma_Z\right)^2} \left[\frac{\left(s + M_W^2\right)^2 }{s^2} \log \left(\frac{s + M_W^2}{M_W^2}\right) - \frac{M_W^2}{s} - \frac{3}{2}\right]\Bigg\}.
\end{align} 

Nevertheless, we will continue using Eq.~\eqref{eq:reducedCrossSec}, since the introduction of the non-unitarity effects can be made in a more transparent way. 

From Eq.~\eqref{eq:differential}, the total cross section is	
\begin{equation}
\sigma(s) = \int_{x_\mathrm{min}}^{1} dx \int_{-\cos\theta_\mathrm{min}}^{\cos\theta_\mathrm{min}} dy H(x,y;s)\, \sigma_0(s(1 -x)).
\label{eq:NU-totalXS}
\end{equation}

If we now examine this process in a non-unitary mixing framework, it is
almost straightforward to obtain the non-unitarity effects in the reduced
cross section:
	\begin{align}
	\sigma^{NU}_0(s) &= \sum_{i,j}|N_{ei}|^2|N_{ej}|^2 \; \sigma_W(s) F_W(s/M_W^2) \nonumber \\
	&+ \sum_{\alpha,\beta}|(NN^\dagger)_{\alpha\beta}|^2 \; \sigma_Z(s) \nonumber \\
	&+ \sum_{i,j}|N_{ei}|^2|N_{ej}|^2  \; \sigma_{W-Z}(s)F_{W-Z}(s/M_W^2).
	\label{eq:nonUnitaryCross}
	\end{align}
	These corrections can be seen in Fig.~\ref{fig:diagrams}: for
        the $W$ contribution (a), each neutrino line
        contributes with a term $U_{e i}$ in the scattering amplitude,
        while for the $Z$ contribution (b), the provided
        correction is of the form $U_{\alpha i}$. Since the mixing is
        non-unitary, flavor-changing neutral currents are allowed,
        hence the sum must be given over different flavors in the
        second term of Eq.~\eqref{eq:nonUnitaryCross}.
	
	Writing Eq.~(\ref{eq:nonUnitaryCross}) explicitly, we will have 
	\begin{align}
	\sigma^{NU}_0(s) = \frac{G_F^2 s}{12\pi} &\left[ 2\sum_{i,j}|N_{ei}|^2|N_{ej}|^2 \right.\nonumber \\
	&+ \sum_{\alpha,\beta}|(NN^\dagger)_{\alpha\beta}|^2\frac{ (g_V^2 + g_A^2)}{(1 - s / M_Z^2)^2 + \Gamma_Z^2 / M_Z^2} \nonumber \\
	&\left. + \sum_{i,j}|N_{ei}|^2|N_{ej}|^2\frac{2 (g_V + g_A) (1
          - s / M_Z^2)}{(1 - s / M_Z^2)^2 + \Gamma_Z^2 / M_Z^2}
          \right].
        \label{eq:nonUnitaryCross_II} 
	\end{align}
Additionally, as discussed in the previous subsection, there will be NU
corrections to $G_F$ and $\sin^2\theta_W$ as described in
Eqs.~(\ref{GfNU}) and~(\ref{weakAngle}) respectively.  Finally, it
should be noticed that in the last two terms of
Eq.~(\ref{eq:nonUnitaryCross}), the decay width, $\Gamma_Z$, appears
in the denominator. Since we are considering the NU case, we must also
introduce the corresponding corrections.  The total $Z$ decay width can
be calculated as~\cite{Akhmedov:1999uz, Tanabashi:2018oca}
	\begin{equation}
	\Gamma_Z = \Gamma_\mathrm{inv} + \Gamma_{\ell\ell} + \Gamma_\mathrm{had}
	\end{equation}
and the non-unitary correction will appear through the
$\Gamma_\mathrm{inv}$ contribution, as it had been computed in the
previous subsection, and we will have:
\begin{equation}
	\Gamma_Z = \frac{G_F
          M_Z^3}{12\sqrt{2}\pi}\sum_{\alpha,\beta}|(NN^\dagger)_{\alpha\beta}|^2
        + \Gamma_{\ell\ell} + \Gamma_\mathrm{had} .
\end{equation}
	
Now that we have introduced the theoretical expressions for the two
neutrino counting observables with the formalism for the non-unitary
case, in the triangular parameterization, we will discuss the
corresponding current constraints and future perspectives for these
two cases.

\section{Experimental tests}

\subsection{The process $e^-e^+ \to \nu\bar{\nu}\gamma$}
To obtain constraints for the NU case from the process $e^-e^+ \to
\nu\bar{\nu}\gamma$, we use the reported measurements from the 
ALEPH~\cite{Heister:2002ut}, DELPHI~\cite{Abreu:2000vk},
L3~\cite{Acciarri:1997dq, Acciarri:1998hb, Acciarri:1999kp}, and
OPAL~\cite{Ackerstaff:1997ze, Abbiendi:1998yu, Abbiendi:2000hh}
collaborations. They are are listed in 
Table~\ref{Tab:experimentsAttributes}. The center of mass energy 
for each run is listed in the first column. The background subtracted 
measured and Monte Carlo cross sections are given in columns two and 
three, respectively. The number of observed events after background 
subtraction are given in column four, while the efficiency corresponds 
to column five. Lastly, the kinematical cuts for the outgoing photon 
energy and angle are reported in the last two columns. For these cuts, 
$x_T = x \sin\theta_\gamma$ (with $x=E_\gamma/E_\mathrm{beam}$), 
while $y = \cos\theta_\gamma$.

	\begin{table}[h]
		\centering
		\begin{tabular}{ccc@{\hskip 0.2in}c@{\hskip 0.3in}c@{\hskip 0.3in}c@{\hskip 0.3in}c@{\hskip 0.3in}c@{\hskip 0.3in}c}
			\toprule				
			 & & $\sqrt{s}$ (GeV) & $\sigma^\mathrm{mes}$ (pb) & $\sigma^\mathrm{MC}$ (pb)
			 & $N_\mathrm{obs}$	 & $\epsilon(\%)$ & $E_\gamma$ (GeV) & $|y|$ \\
			\cline{1-9}      
			 \multirow{10}{*}{ALEPH} & \multirow{2}{*}{\cite{Barate:1997ue}} & 161  & $5.3\pm0.83$ & $5.81\pm0.03$ & 41 & 70 & $x_T\geq 0.075$ & $\leq 0.95$ \\
			  & & 172  & $4.7\pm0.83$ & $4.85\pm0.04$ & 36 & 72 & $x_T\geq 0.075$ & $\leq 0.95$ \\
			  \cline{2-9}
			  & \cite{Barate:1998ci} & 183  & $4.32\pm0.34$ & $4.15\pm0.03$ & 195 & 77 & $x_T\geq 0.075$ & $\leq 0.95$ \\
			  \cline{2-9}
              & \multirow{7}{*}{\cite{Heister:2002ut}} & 189  & $3.43\pm 0.17$ & $3.48\pm 0.05$ & 484 & \multirow{8}{*}{81.5} & 
                     \multirow{8}{*}{$x_T \geq 0.075$} & \multirow{8}{*}{$\leq 0.95$}\\
			  		 & & 192  & $3.47\pm 0.40$ & $3.23\pm 0.05$ & 81  & & & \\
			  		 & & 196  & $3.03\pm 0.23$ & $3.26\pm 0.05$ & 197 & & & \\
			  		 & & 200  & $3.23\pm 0.22$ & $3.12\pm 0.05$ & 231 & & & \\
			  		 & & 202  & $2.99\pm 0.29$ & $3.07\pm 0.05$ & 110 & & & \\
			  		 & & 205  & $2.84\pm 0.22$ & $2.93\pm 0.05$ & 182 & & & \\
			  		 & & 207  & $2.67\pm 0.17$ & $2.80\pm 0.05$ & 292 & & & \\
			\cline{1-9}
			\multirow{3}{*}{DELPHI} & \multirow{3}{*}{\cite{Abreu:2000vk}}	& 189  & $1.80\pm 0.20$  & $1.97$  & 146 & 51 & $x\geq 0.06$ & $\leq 0.7$ \\
					& & 183  & $2.33\pm 0.36$  & $2.08$  & 65 & 54 & $x\geq 0.02$ & $\leq 0.85$ \\
					& & 189  & $1.89\pm 0.22$  & $1.94$  & 155 & 50 & $x\leq 0.9$	& $\leq 0.98$ \\
			\cline{1-9}
			\multirow{6}{*}{L3}	& \multirow{3}{*}{\cite{Acciarri:1997dq}}	& 161  & $6.75\pm 0.93$  & $6.26\pm 0.12$  & 57 & 80.5 & $\geq 10$ &  $\leq 0.73$  \\
			& & & & & & & and & and \\
					& & 172  & $6.12\pm 0.90$  & $5.61\pm 0.10$  & 49 & 80.7 & $E_T\geq 6$ & 0.80-0.97 \\
			\cline{2-9}		
					& \cite{Acciarri:1998hb} & 183  & $5.36\pm 0.40$  & $5.62\pm 0.10$  & 195 & 65.4 &$\geq 5$ &  $\leq 0.73$  \\
					& & & & & & & and & and \\
					& \cite{Acciarri:1999kp} & 189  & $5.25\pm 0.23$  & $5.29\pm 0.06$  & 572 & 60.8 &  $E_T\geq 5$ & 0.80-0.97 \\
			\cline{1-9}
			\multirow{10}{*}{OPAL} & \multirow{3}{*}{\cite{Ackerstaff:1997ze}}	& 130  & $10.0\pm 2.34$  & $13.48\pm0.22$ & 19 & 81.6 & $x_T > 0.05$ &  $\leq 0.82$  \\
			& & & & & & & and & and \\
					& & 136  & $16.3\pm 2.89$  & $11.30\pm 0.20$  & 34 & 79.7  & $x_T > 0.1$ &  $\leq 0.966$ \\
			\cline{2-9}
					& \multirow{2}{*}{\cite{Abbiendi:1998yu}} & 130  & $11.6\pm 2.53$  & $14.26\pm 0.06$  & 21 & 77 & \multirow{2}{*}{$x_T > 0.05$} &  \multirow{2}{*}{$\leq 0.966$}\\
					& & 136  & $14.9\pm 2.45$  & $11.95\pm 0.07$  & 39 & 77.5 &  &  \\
			\cline{2-9}
					& \multirow{3}{*}{\cite{Ackerstaff:1997ze}} & 161  & $5.30\pm 0.83$  & $6.49\pm 0.08$  & 40 & 75.2 & $x_T > 0.05$ &  $\leq 0.82$  \\
					& & & & & & & and & and \\
					& & 172  & $5.50\pm 0.83$  & $5.53\pm 0.08$  & 45 & 77.9 & $x_T > 0.1$ & $\leq 0.966$\\
			\cline{2-9}
					& \cite{Abbiendi:1998yu} & 183  & $4.71\pm 0.38$  & $4.98\pm 0.02$  & 191 & 74.2 & $x_T > 0.05$ & $\leq 0.966$\\
			\cline{2-9}
					& \cite{Abbiendi:2000hh} & 189  & $4.35\pm 0.19$  & $4.66\pm 0.03$  & 643 & 82.1 & $x_T > 0.05$ & $\leq 0.966$ \\			          
			\botrule
\end{tabular}
\caption{Summary from the ALEPH~\cite{Barate:1997ue, Barate:1998ci, Heister:2002ut}, DELPHI~\cite{Abreu:2000vk},
	L3~\cite{Acciarri:1997dq, Acciarri:1998hb, Acciarri:1999kp}, and
	OPAL~\cite{Ackerstaff:1997ze, Abbiendi:1998yu, Abbiendi:2000hh} collaboration experimental data, collected above the $W^+W^-$ production threshold.}
\label{Tab:experimentsAttributes}
\end{table}

In order to make our analysis, we have computed the cross section from
Eqs.~(\ref{eq:NU-totalXS}) and~(\ref{eq:nonUnitaryCross_II}), with the
integration limits taken according to the last two columns of
Table~\ref{Tab:experimentsAttributes}. Although we have a good
  agreement in our integration with many of the reported Monte Carlo
  simulations, there are some exceptions due, we believe, to our lack
  of knowledge of each experimental setup. Instead of excluding any
  experimental value, we have included a normalization error in our
  analysis, with a $10$~\% uncertainty.
Once we have obtained this expression, we have compared our theoretical
expectation for the NU case with the experimental results of
Table~\ref{Tab:experimentsAttributes} through a $\chi^2$ analysis.

\begin{figure}
		\centering 
		\includegraphics[scale=0.6]{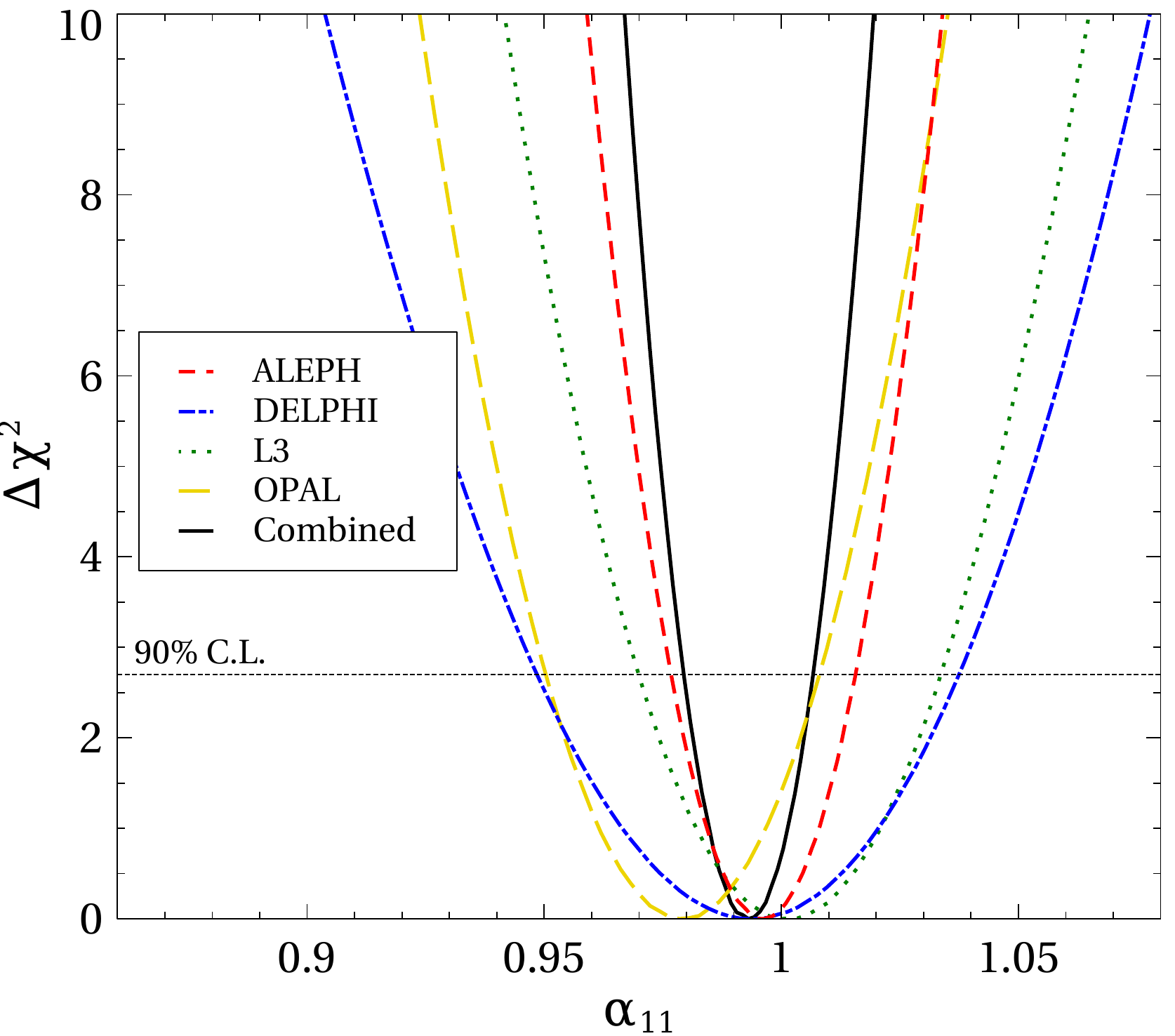}
		\caption{Bounds on the NU parameter $\alpha_{11}$ from the 
                  process $e^-e^+ \to \nu\bar{\nu}\gamma$, using the
                  ALEPH, DELPHI, L3, and OPAL reported results.}
		\label{fig:ALEPHbest}
	\end{figure}

 Our result for the non-unitary parameter
  $\alpha_{11}$ is shown in Fig.~\ref{fig:ALEPHbest}, for each experiment
  separately, and for a combination of all of them. In this
  analysis, we have considered any other NU parameter as equal to the
  Standard case, that is, $\alpha^{2}_{22} = \alpha^{2}_{33} = 1$ and
  $\alpha^{2}_{21} = \alpha^{2}_{31} =\alpha^{2}_{32} = 0$.  We have
  chosen this parameter because diagonal parameters $\alpha_{ii}$
  give the main contribution for deviations from unitarity. Besides,
  any diagonal parameter contributes on equal footing and, therefore,
  our constrain can be equally applied to $\alpha_{22}$ or
  $\alpha_{33}$. As it can be seen, it is possible to restrict the
$\alpha_{11}$ NU parameter, and the constraint at $90$~\% CL is given
by
\begin{equation}
\alpha_{11} > 0.9794, \; \; \;  1 - \alpha_{11} < 0.0206.
\end{equation}
To our knowledge, this is the first time that a constraint for NU is
reported using this observable and it is possible to see that the
limits are competitive. 

\subsection{The invisible Z decay}

We now turn our attention to the particular case of the $Z$ decay into
neutrinos. This process has already been measured by
LEP~\cite{ALEPH:2005ab, Schael:2013ita} and future
experiments~\cite{Fan:2014vta, Baak:2013fwa, Baer:2013cma,
    Banerjee:2015gca, Gomez-Ceballos:2013zzn, Blondel:2014bra, CEPCStudyGroup:2018ghi,CEPC-SPPCStudyGroup:2015csa,
    Liao:2017jiz, Liang:2018mst} can improve the
measurement of this important observable. Previous works have already
reported constraints on NU parameters using this observable for a
combined analysis from different
measurements~\cite{Antusch:2014woa,
    Antusch:2015mia,Antusch:2017hvv, Fernandez-Martinez:2016lgt}. Here
  we focus on
this particular parameter using the specific triangular parameterization and 
making more emphasis in the perspectives from future experimental proposals. 

\begin{figure}[b]
\begin{center}
  \includegraphics[width=.6\textwidth,angle=0]{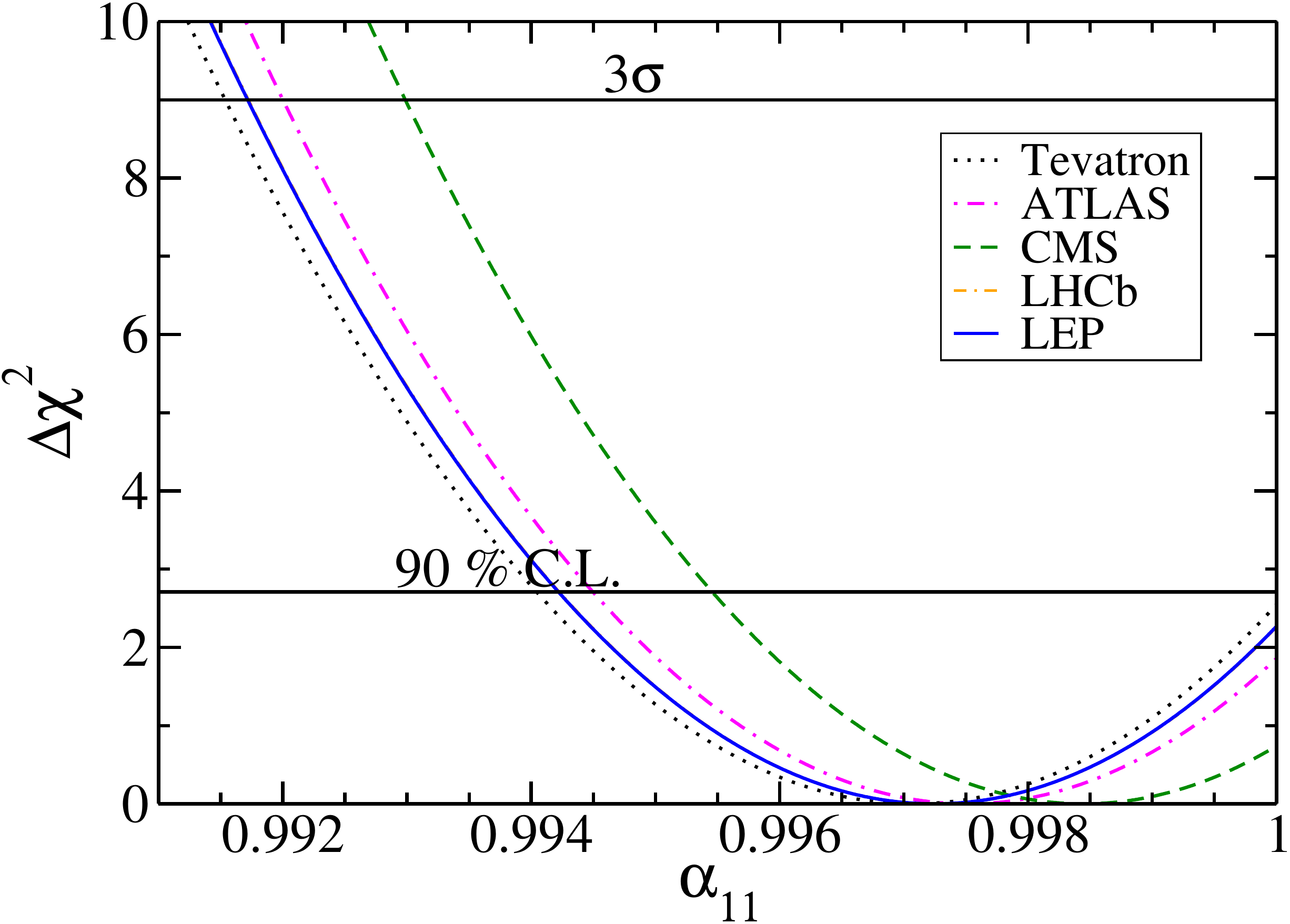}
  \caption{Restrictions for $\alpha_{11}$ from the invisible decay of
    the $Z$ boson, depending on the value of the effective weak mixing
    angle, $\bar{s}^2_l$. We consider the measurements on
    $\bar{s}^2_l$ coming from different experiments.  }
  \label{fig:a11_chi2}
\end{center}
\end{figure}

Before analyzing the invisible decay constraints on NU, it is
important to remember from the previous section that the NU case will
affect the theoretical prediction of different parameters, such as
$G_F$ and $\sin^2\theta_W$ (Eqs.~\eqref{GfNU}
  and~\eqref{weakAngle} respectively.)  Perhaps the most important
observable for our discussion is the value of the weak mixing angle
that, at the relevant energy, differs up to three standard deviations
depending on the experiment that measures it. Its
  impact is illustrated in Fig.~\ref{fig:a11_chi2}, where we show the
  $\chi^2$ curve for this observable as a function of the
  $\alpha_{11}$ parameter.  In this figure, besides considering the
  LEP~\cite{ALEPH:2005ab, Schael:2013ita} measurement for the weak
  mixing angle, we also show how this constraint changes if we
  consider other measurements for the weak mixing angle. That is the
case of the
Tevatron~\cite{Abazov:2011ws,Acosta:2004wq,Aaltonen:2013iut,TEW:2010aj},
Atlas~\cite{Aad:2015uau}, LHCb~\cite{Aaij:2015lka} and
CMS~\cite{Chatrchyan:2011ya} result. It is possible to notice that the
evaluation of this fundamental quantity of the Standard Model still
can have an impact on the non-unitarity constraints. 
  As in the previous subsection, for this plot we have only considered
  $\alpha_{11}$ as different from one and all other non-unitary
  parameters as equal to the standard case, that is, $\alpha^{2}_{22} =
  \alpha^{2}_{33} = 1$ and $\alpha^{2}_{21} = \alpha^{2}_{31}
  =\alpha^{2}_{32} = 0$.

Provided that we have a precise measurement of the weak mixing angle,
we can return to the computation of constraints on NU from current
and future experimental proposals that will improve the measurements of
different observables, such as the number of neutrinos, $N_{\nu}$, or
the effective value of the weak mixing angle,
$\sin^2{\theta_{eff}}$. We show their sensitivity  in Table~\ref{table:future_exp_val}.
\begin{table} [h]
 \begin{tabular}{lcccc}
\hline \hline & \quad LEP~\cite{ALEPH:2005ab} & \quad CEPC~\cite{CEPC-SPPCStudyGroup:2015csa}&\quad FCC-ee~\cite{Gomez-Ceballos:2013zzn}&\quad ILC~\cite{Baer:2013cma}  \\
\hline \hline
$\sigma{(N_{\nu})}[10^{-3}]$&\quad$8.0$&\quad$3.0$&\quad$4.0$&\quad$4.0$\\
\hline
$\sigma{(\sin^2{\theta_{eff}})}[10^{-4}]$&\quad$1.6$&\quad$0.23$&\quad$
                                                                 0.01$&\quad$0.1$\\
\hline \hline 
\end{tabular}
\caption{Expected uncertainties on $N_{\nu}$ and
  $\sin^2{\theta_{eff}}$ for different experimental proposals. Notice
  that for LEP we quote the present experimental values, whereas for
  CEPC, FCC-ee, and ILC we show future estimations.}
\label{table:future_exp_val}
\end{table}

\begin{table}[h!]
 \begin{tabular}{lccc}
\hline \hline &&CEPC \\
\hline
$R^0_{inv}$&$\;5.9430 \pm 0.0065$&\quad $5.9671 \pm 0.0065$ \quad &\;$5.9801 \pm 0.0065$\\
\hline &&FCC-ee / ILC \\
\hline
$R^0_{inv}$&$ \;5.9430 \pm 0.0083$&\quad $ 5.9671 \pm 0.0083$ \quad &\;$ 5.9801 \pm 0.0083$\\
\hline \hline 
\end{tabular}
 \caption{Test values for the invisible ratio $R^0_{inv}$ used in the present work. We quote the expected uncertainty
   coming from future experiments. }
\label{table:R_future}
\end{table}

In order to estimate the sensitivity of the future experiments we
will consider again the ratio given by Eq.~(\ref{ratioExp}).
In particular, the uncertainty of $R^0_{inv}$ is calculated from
\begin{equation}
\sigma^2(R^0_{inv}) = \left(\frac{\Gamma_{\nu \bar{\nu}}}{\Gamma_{l \bar{l}}}\right)^2_{SM}
\sigma^2(N_{\nu}) + (N_{\nu})^2 \; \sigma^{2}\left(\frac{\Gamma_{\nu \bar{\nu}}}{\Gamma_{l \bar{l}}}\right)_{SM}\, ,
\nonumber
\end{equation} 
where $\sigma\left(\frac{\Gamma_{\nu \bar{\nu}}}{\Gamma_{l \bar{l}}}\right)_{SM}=\;
0.00083$~\cite{ALEPH:2005ab} and $\sigma(N_{\nu})$ is given in
Table~\ref{table:future_exp_val}. With these hypothesis we 
obtain the results shown in Table~\ref{table:R_future}.

Within this framework, it is possible to obtain an idea of the future
sensitivity of these experiments on the NU parameters. A forecast for
this sensitivity can be computed considering three different cases of
a future measurement of the ratio $R^0_{inv}$:
\begin{itemize}
 \item The experimental value reported at~\cite{ALEPH:2005ab}, $R^0_{inv}
   = 5.9430$.  
 \item The theoretical value calculated from the effective weak mixing angle
   including radiative corrections~\cite{Patrignani:2016xqp}, $R^0_{inv}
   = 5.9671$.
 \item A value two standard deviations (of CEPC) above of the previous value, $R^0_{inv}
   = 5.9801$.
\end{itemize}

\begin{figure}
\begin{center}
\includegraphics[width=.5\textwidth,angle=0]{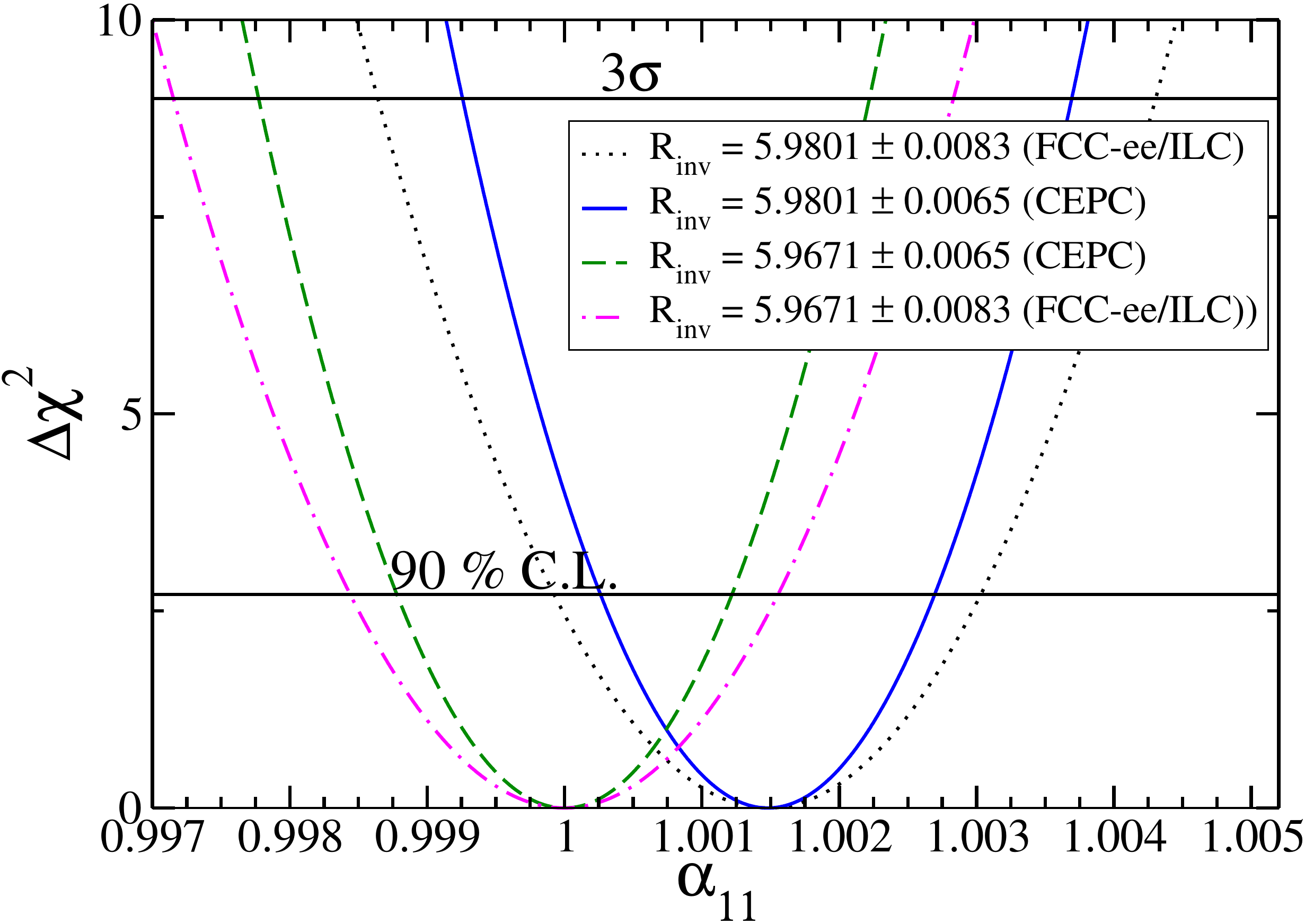}
\includegraphics[width=.465\textwidth,angle=0]{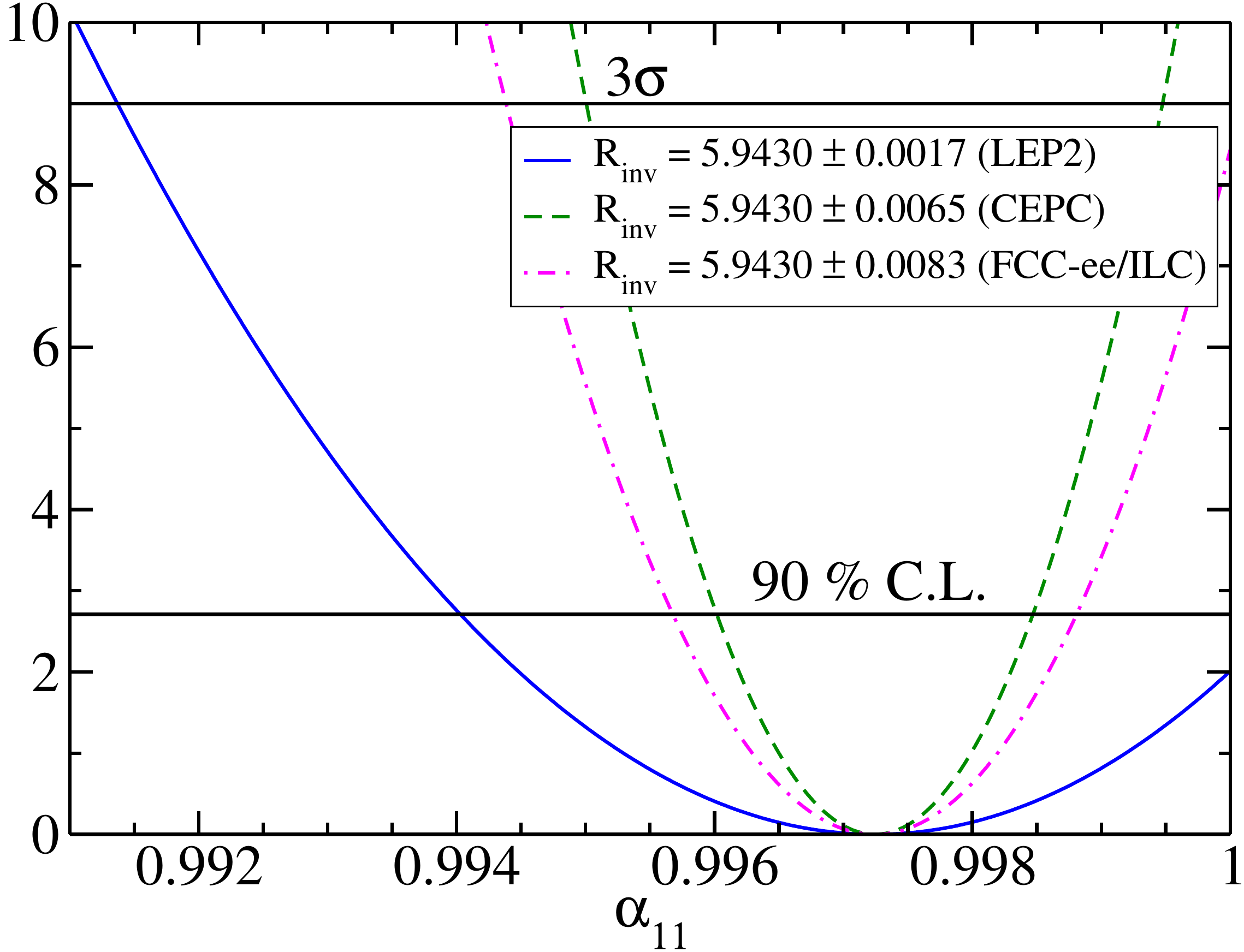}
  \caption{Restrictions for $\alpha_{11}$ from the invisible decay of
    the $Z$ boson, for the future proposals
    CEPC, FCC-ee and ILC experiments. We have considered different
    possible central values to illustrate the constraints to be
    obtained.}
    \label{fig:chi_a11}
\end{center}
\end{figure}

\begin{figure}
\begin{center}
\includegraphics[width=.5\textwidth,angle=0]{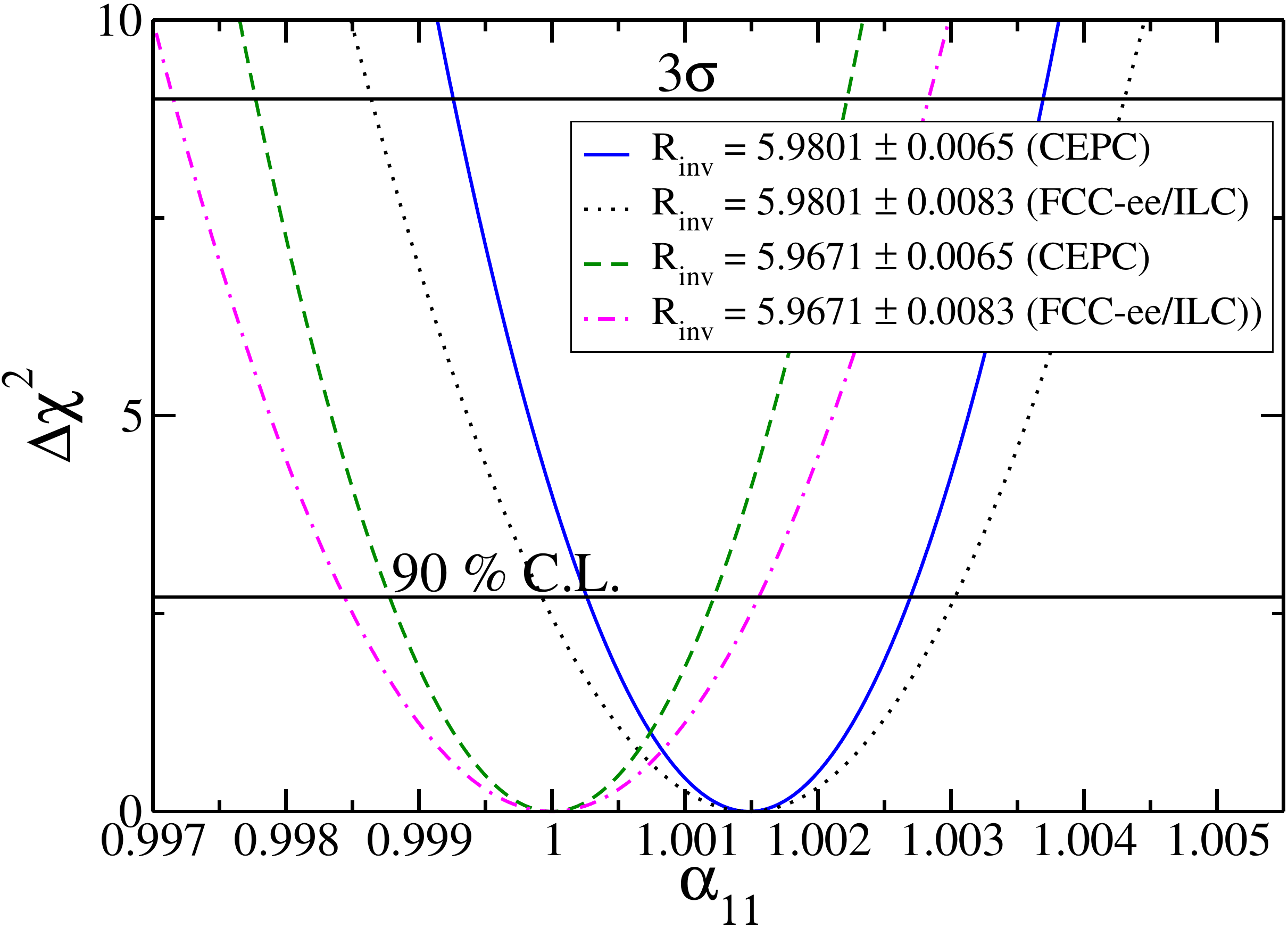}
\includegraphics[width=.465\textwidth,angle=0]{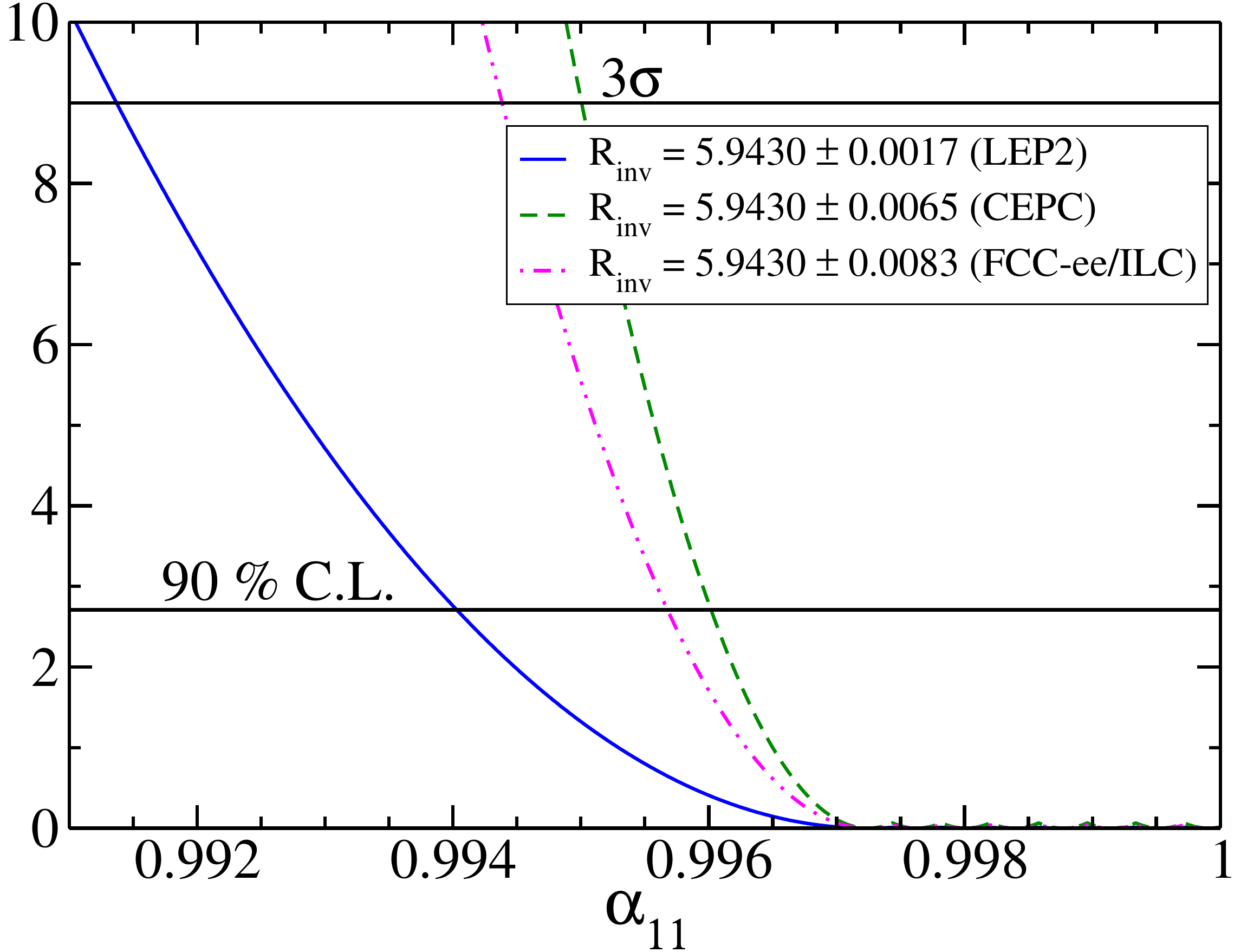}
  \caption{Restrictions for $\alpha_{11}$ from the invisible decay of
    the $Z$ boson for the  future CEPC, FCC-ee and ILC
    experiments. Different central values have been used as a test to
    illustrate the possible constraints. For this case,
    we have considered $\alpha_{33}$ and $\alpha_{31}$ as free
    parameters in the fit (fulfilling the Cauchy-Schwartz condition).}
\label{fig:chi_a11_a31CS}
\end{center}
\end{figure}

  To consider these futuristic scenarios, it takes into
  account the possible non-standard result where the effective number
  of neutrinos is smaller than three. Besides, it also considers the
  less expected case where a future experiment might have a
  statistical fluctuation, and measures a value above the SM
  prediction. For these three cases, we perform a $\chi^2$ analysis
in order to have a forecast of the future expected sensitivity,
considering the following two scenarios:
\begin{itemize}
 \item Firstly, we consider that $\alpha_{11}$ is the only parameter
different from the standard case. The $\chi^2$ fit is made with the
errors already discussed for each experiment. The results are compiled in Fig.~\ref{fig:chi_a11}.
 \item  Secondly, we let $\alpha_{11}$, $\alpha_{13}$
   and $\alpha_{33}$ to vary freely, while fulfilling the Cauchy-Schwarz
   condition:
\begin{equation}
|\alpha_{ij}| \leq \sqrt{(1-\alpha_{ii}^{2}) \, (1-\alpha_{jj}^{2})}
\, .
\label{eq:Cau_Sch}
\end{equation} 
The other NU parameters are set to their SM value. The results
obtained are shown in Fig.\ref{fig:chi_a11_a31CS}. Notice that we have
considered only $\alpha_{33}$ and $\alpha_{31}$ different from zero,
since very similar results will be obtained with $\alpha_{22}$ and
$\alpha_{21}$.
\end{itemize} 
We summarize the expected accuracy for both cases in
Table~\ref{table:chi_a11_res}.  We can see from these results that
future collider experiments could give a constraint on the diagonal
non-unitary parameter that will be stronger than the current global
limits~\cite{Fernandez-Martinez:2016lgt,
  Blennow:2016jkn,Escrihuela:2016ube}, that constraints $\alpha_{11}$
at the level of $0.999$ or below as we see in
Table~\ref{table:curr_bounds}. It is also interesting
  to notice what would be the constraint in the case of a measurement
  different from the SM prediction; as illustrated in the same
  Table~\ref{table:chi_a11_res} the future experiments under
  discussion will have the potential to show the evidence of new
  physics through a non-unitarity of the neutrino mixing-matrix. 

\begin{table}
 \begin{tabular}{lccc}
\hline \hline Experiment &$R^{0}_{inv}$ &$\alpha_{11}$  & $\alpha_{11}$  ($\alpha_{31}$ and $\alpha_{33}$ free) \\
\hline \hline
current  &$5.9430$ &$0.99403 \; < \; \alpha_{11}$& $0.99403 \; < \; \alpha_{11}$\\
\hline
\hline
CEPC & $5.9430$  & \quad$0.99602 \; < \; \alpha_{11}  \;<\; 0.99847$ & 
$0.99602 \; < \; \alpha_{11}$\\

\hline
FCC-ee/ILC & $5.9430$ & \quad$0.99568 \; < \; \alpha_{11}
                                 \;<\; 0.99881$ & $0.99568 \; <
                                                  \; \alpha_{11} $\\

\hline
CEPC & $5.9671$ & $0.99879 \; < \;
                                     \alpha_{11}\; < \;
                  1.00123$&$0.99879 \; < \; \alpha_{11}\; < \;
                            1.00123$\\

\hline
FCC-ee/ILC & $5.9671$ & $0.99844 \; < \;
                                    \alpha_{11}\; < \; 1.00156$&$0.99845 \; < \;
                                    \alpha_{11} \; < \; 1.00157$\\

\hline
CEPC & $5.9801$ & $1.00026 \; < \;
                                     \alpha_{11}\; < \;
                  1.00269$&$1.00027 \; < \; \alpha_{11} \; < \; 1.00270$\\

\hline
FCC-ee/ILC & $5.9801$ & $0.99993 \; < \;
                                    \alpha_{11}; < \; 1.00305$&$0.99994 \; < \;
                                    \alpha_{11} \; < \; 1.00304$\\

\hline \hline
\end{tabular}
\caption{Allowed values for $\alpha_{11}$ at $90$~\% 
  C.L., considering present experimental values and future proposals
  from CEPC, FCC-ee and ILC experiments. We consider either the case
  when any NU parameter other than $\alpha_{11}$ is in the unitary limit 
  and also when   $\alpha_{31}$ and $\alpha_{33}$ are allowed
  to vary, fulfilling the Cauchy-Schwartz condition.}
\label{table:chi_a11_res}

\end{table}

\begin{table}
 \begin{tabular}{cc|c}
\hline \hline 
\multicolumn{3}{c}{Current bounds}\\
\hline \hline
\multicolumn{2}{c|}{$\alpha_{11}\; (90~\%
   \,\text{C.L.})$~\cite{Escrihuela:2016ube}} &$\alpha_{11}\; (2\sigma)$~\cite{Fernandez-Martinez:2016lgt, Blennow:2016jkn} \\
\hline
One parameter & All parameters&  \\

$\alpha_{11} >0.98000$ & $\alpha_{11} >0.96000$& $\alpha_{11} >0.99875$ \\

\hline \hline
\end{tabular}
\caption{ Current bounds on non-unitary $\alpha_{11}$ parameters coming
from~\cite{Fernandez-Martinez:2016lgt, Blennow:2016jkn,Escrihuela:2016ube}.}
\label{table:curr_bounds}
\end{table}

\section{conclusions}
We have reviewed the measurements for neutrino counting observables
close to the $Z$ peak and reported a new analysis for the non-unitary
formalism for the case of the $\nu\bar{\nu}\gamma$ process. The
corresponding constraints have been introduced in this work and we
have shown that they are competitive with other current
constraints. As far as we know, this is the first time this analysis
is done. We have used the triangular parameterization to perform this
analysis.

We have also analyzed the invisible $Z$ decay into neutrinos, in the
same triangular parameterization. In this case we have focused in the
importance of a precise determination of the weak mixing angle and in
the perspectives to improve current constraints by using future
collider experiments, that are expected to be constructed as a
continuation of the precision program for particle physics. They will
allow to obtain better restrictions to new physics from several
processes at different energy regimes.  For this purpose, we have
focused in the invisible decay width in the $Z$ peak, that will be
measured in the first stages of the future collider experiments ILC,
FCC-ee and CEPC.

We have shown that any of these experiments will have enough
sensitivity to improve the current constraint on non-unitarity. We
have focused especially in the diagonal parameter $\alpha_{11}$. To
obtain this result we have used different test values. In particular,
for a measurement as low as the current LEP central value, future
experiments will give a positive signal for non-unitarity at $90$ \%
C. L., while a future measurement in accordance with the Standard model
prediction will restrict the limit for $\alpha_{11}$ to be bigger
that $0.999$, that is, a precision at the level of $10^{-3}$. It is
also important to notice that, as can be seen from Eq.~(\ref{eq:NN}), the
$Z$ decay measurement will mainly restrict the sum of the three
diagonal parameters: $\sum_i \alpha_{ii}$ and, therefore, in a combined 
analysis, this measurement will help to restrict any of the diagonal 
parameters.

\section{Acknowledgments}
This work was supported by the Conacyt grant A1-S-23238 and SNI
(Mexico). LJF also thanks the Conacyt for the grant of Ayudante de
investigador (EXP. AYTE. 16959) and a posdoctoral CONACYT grant.

\bibliographystyle{apsrev} 
\providecommand{\url}[1]{\texttt{#1}}
\providecommand{\urlprefix}{URL }
\providecommand{\eprint}[2][]{\url{#2}}


\end{document}